\documentclass{PoS}

\newcommand{\bra}[1]{\langle #1|}
\newcommand{\ket}[1]{|#1\rangle}
\newcommand{\bkt}[1]{\langle #1\rangle}
\newcommand{\ba}{\begin{align}}
\newcommand{\ea}{\end{align}}
\newcommand{\beq}{\begin{eqnarray}}
\newcommand{\eeq}{\end{eqnarray}}

\title{Proper heavy-quark potential from
 a spectral decomposition of  the thermal Wilson loop}

\ShortTitle{Proper heavy-quark potential in a spectral decomposition from the thermal Wilson loop}

\author{\speaker{A. Rothkopf}\\
        Department of Physics, The University of Tokyo, Tokyo 113-0033, Japan\\
        E-mail: \email{rothkopf@nt.phys.s.u-tokyo.ac.jp}}

\author{T. Hatsuda\\
        Department of Physics, The University of Tokyo, Tokyo 113-0033, Japan\\
        E-mail: \email{hatsuda@phys.s.u-tokyo.ac.jp}}

\author{S. Sasaki\\
        Department of Physics, The University of Tokyo, Tokyo 113-0033, Japan\\
        E-mail: \email{ssasaki@phys.s.u-tokyo.ac.jp}}

\abstract{We propose a 
non-perturbative and gauge invariant derivation of the static potential 
 between
a heavy-quark ($Q$) and an anti-quark ($\bar{Q}$) at finite temperature.
 This {\em proper}  potential is defined through the spectral function (SPF) of the 
  thermal Wilson loop and can be shown to satisfy the Schr\"{o}dinger equation
  for the heavy $Q\bar{Q}$ pair in the thermal medium. In general, 
 the proper potential has a real and an imaginary part,
  corresponding to the peak position and width of the SPF.
 The validity of using a Schr\"{o}dinger equation for  
  heavy  $Q\bar{Q}$  can also be checked from the structure of the SPF.
 To test this idea, quenched QCD simulations on anisotropic lattices
  ($a_\sigma=4a_\tau=0.039\rm fm$, $N^3_\sigma \times N_{\tau} =20^2 \times (96-32)$)
   are performed. The real part of the proper potential below the 
 deconfinement temperature ($T=0.78T_c$) exhibits the well known Coulombic and confining behavior.
 At ($T=2.33T_c$) we find that it  coincides with the Debye screened potential obtained 
  from Polyakov-line correlations in the color-singlet channel
   under Coulomb gauge fixing.
 The physical meaning of the spectral structure of the thermal Wilson loop
  and the use of the maximum entropy method (MEM) 
  to extract the real and imaginary part of the proper potential 
  are also discussed.}

\FullConference{The XXVII International Symposium on Lattice Field Theory\\
		 July 26-31, 2009\\
		 Peking University, Beijing, China}

\begin{document}

\section{Introduction}
Heavy quarkonium at finite temperature is a both intriguing and challenging subject.
 In particular, its in-medium behavior has 
been proposed to be a prime signal for the creation of the quark-gluon plasma (QGP),
 expected above the critical temperature $T_c$ of the deconfinement
 transition \cite{Matsui:1986dk}.
 Moreover, the advent of RHIC has made it possible to access  
  the QGP in the laboratory \cite{Yagi:2005yb}
  and a partial suppression of charmonium
   has been experimentally confirmed \cite{Rapp:2009my}.

 In the meantime, theoretical understanding has also progressed significantly. 
 Originally, a non-relativistic Schr\"{o}dinger equation 
 with screened Coulomb potential was considered for the heavy $Q\bar{Q}$
 pair above $T_c$ \cite{Matsui:1986dk}. 
 Later on, more sophisticated potentials were adopted, such as 
 the color-singlet free-energy of a static $Q\bar{Q}$ pair and its variants
 \cite{Satz:2008zc}.
  A main drawback of these approaches is that
  there is no firm theoretical foundation to use such
  potentials within the Schr\"{o}dinger equation of a heavy $Q\bar{Q}$ pair
  in a hot environment.\footnote{This  is in contrast to the situation at zero temperature in which  
   a systematic way to derive  the potential to be used in the 
   Schr\"{o}dinger equation has been formulated
   in the framework of non-relativistic QCD  \cite{Brambilla:2004jw}.} A more direct approach to heavy quarkonium at finite temperature $T$ is to 
 extract the spectral function (SPF) from lattice QCD simulations 
 with the help of the maximum entropy method (MEM)
  \cite{Asakawa:2000tr}.
  An unexpected feature found subsequently was that charmonium may survive 
  even above the deconfinement transition up to about 2$T_c$  \cite{Asakawa:2003re,DeTar:2009ef}. 
  However, a transparent understanding of this result has not been obtained so far. 
    
   In light of these circumstances it is imperative to establish a solid connection between
   the Schr\"{o}dinger approach and the spectral function approach, based on a proper 
   definition of the in-medium potential.
  A possible way to reach this goal was initially proposed 
   in \cite{Laine:2006ns};
   the proper potential to be used in the 
    in-medium Schr\"{o}dinger equation was defined from the late-time ($t$)
    behavior of the forward correlator $D^{>}(t,R)$ of a heavy $Q\bar{Q}$ pair
     separated by distance ($R$). It was shown by using hard thermal resummation (HTL) techniques 
     at high $T$ that the potential has both a real and an imaginary part. 
  The purpose of the present article is to develop this idea further by
  introducing a spectral decomposition of $D^{>}(t,R)$ (or equivalently the
  thermal Wilson loop) and to explore the non-perturbative 
  derivation of the real and imaginary part of the 
  proper $Q\bar{Q}$ potential on the basis of lattice QCD simulations.
  If this program turns out to be 
  successful, a true physical understanding of the 
  heavy-quark bound state in the QGP can possibly be obtained.

\section{Formulation}

\subsection{Spectral function for a static $Q\bar{Q}$ pair}
\label{sec:SPF}

We start with the  $Q\bar{Q}$ operator $M_R(t)$ defined as
$ M_R(t)=\bar{\psi}(x)\Gamma U_P(x,y)\psi(y)$, 
 where $R=|\mathbf{x}-\mathbf{y}|$ and $t=x^0=y^0$. The Wilson-line operator
 $U_P(x,y)$ is chosen to connect the points $x=(t,{\bf x})$ and $y=(t,{\bf y})$
 by a straight space-like path $P$. An arbitrary 
 Dirac matrix is denoted  by $\Gamma$,  e.g. $\Gamma=\gamma^\mu$ in the vector channel and 
 $\Gamma=i\gamma_5$ in the pseudoscalar channel. Let us introduce the 
 forward correlation function of the $Q\bar{Q}$ pair,
 \beq 
 D^>(t,R)= \bkt{M_R(t)M_R^\dagger(0)} \ \ \ \ \ \ \ (t >0)
 \label{DefD}.
\eeq 
  In a fully dynamic setting, heavy quarks can propagate in time, 
 hence the relative distance between $Q$ and $\bar{Q}$ will change accordingly.
 On the other hand, if we consider the infinite mass case 
 ($m_Q \rightarrow \infty$), the spatial separation $R$ 
 reduces to an external parameter.
 The spectral function associated with Eq.(\ref{DefD}) reads
\vspace{-0.17cm}
\beq
\! \! \! \! \! \! \rho(\omega,R)=(1- e^{-\beta \omega } ) \tilde{D}^>(\omega,R)
 =\frac{1}{Z}\sum_{n,n'} |\bra{n}M_R(0)\ket{n'}|^2 \Big(e^{-\beta E_n}
 -e^{-\beta E_{n'}}\Big)
 \delta(\omega-(E_{n'}-E_n)),
\label{eq:SPF}
\eeq
where $\beta=1/T$, $\tilde{D}^>(\omega,R)$ is the Fourier transform of ${D}^>(t,R)$, and 
 $Z$ is the full partition function of the system with gluons, light quarks and 
 heavy quarks.   Eq.(\ref{eq:SPF}) is antisymmetric in $\omega$ due to the 
 bosonic character of the operator $M_R$.
 
 We now consider heavy quarks with nearly infinite mass $m_Q \rightarrow \infty$,
  so that they cannot move spatially and their relative distance $R$ stays fixed.
 The spectral function can thus be decomposed into the sum of 
 the three contributions, 
 $\rho(\omega,R)= \rho_{\rm I}(\omega,R)+\rho_{\rm II}(\omega,R)+\rho_{\rm III}(\omega,R)$
 depending on the intermediate states given below ("light" implies light quarks and gluons):
\begin{center}
\begin{tabular}{c|ccc}
\hline
&  I &  II &  III  \\
\hline \hline 
\vspace{-0.37cm}
& & & \\
$ |n'\rangle$ & $Q\bar{Q}\,+\,{\rm light}$ & light  & $Q$ or $\bar{Q}\,+\,{\rm light} $ \\ 
$|n \rangle$ &light  & $Q\bar{Q}\,+\,{\rm light}$ & $\bar{Q}$ or $Q\,+\,{\rm light} $ \\

\hline
\end{tabular}
\end{center}

 Since we are interested in the interaction between $Q$ and $\bar{Q}$ with 
 a spatial separation $R$ at the same point in time,
  only the contributions I and  II are relevant for our purpose.
    Moreover,  I and  II are simply related with each other by
   $\rho_{\bf II}(\omega,R)= - \rho_{\bf I}(-\omega,R)$, due to the      
 anti-symmetric nature of the SPF.
 Without loss of generality, it is therefore sufficient to focus on the contribution I.
  By taking into account the fact that $m_{Q} \gg T$, we 
 obtain
\beq
 \rho_{\rm I}(\bar{\omega},R)=\frac{1}{Z_0}
 \sum_{n,n'} |\bra{n}M_R(0)\ket{n'}|^2 
 \delta(\bar{\omega}-(\epsilon_{n'}(R)-\epsilon_n)) e^{-\beta \epsilon_n},
\label{eq:SPF-I}
\eeq
where $Z_0$ is the partition function without the heavy quarks, $|n\rangle$ and 
$| n' \rangle$ belong to the case I in the above Table.
 Also, we have defined $\bar{\omega} \equiv \omega - 2m_Q$,  
  $\epsilon_{n'} (R) \equiv E_{n'} -2m_Q$
 and $\epsilon_n \equiv E_n$.  Note that $\epsilon_{n'} (R)$ is the 
 $R$-dependent energy of a $Q\bar{Q}$ pair measured from the
  total rest mass $2m_Q$, while  $\epsilon_{n} $ is $R$-independent since no
   heavy quarks are present in the state $|n\rangle$. 
   Note that  they are both $T$-independent by definition.
  
   The spectral structure of 
  $ \rho_{\rm I}(\bar{\omega},R)$  contains all the information of the 
  interaction between the heavy quarks in the hot medium. 
 To see its connection to the Schr\"{o}dinger equation for the heavy quark
  system, let us start with the following relation obtained from the 
  definitions of $D^>_{\rm I}$ and $\rho_{\rm I}$:
\beq
i\partial_t D_{\rm I}^>(t,R)=2m_QD_{\rm I}^>(t,R)+e^{-i2m_Q t}\int_{-\infty}^{\infty}
 e^{-i\bar{\omega}t}\bar{\omega} \rho_{\rm I}(\bar{\omega},R) d\bar{\omega} .
\label{Schroed}
\eeq
  If $\rho_{\rm I}(\bar{\omega},R)$ has a distinct Breit-Wigner peak
    at  $\bar{\omega} = {\omega}(R,T)$ with a half-width $\xi(R,T)$,
  Eq.(\ref{Schroed}) reduces to  the Schr\"{o}dinger equation,
\beq
i\partial_t D_{\rm I}^>(t,R)&=&\Big[ 2m_Q+ {\omega}(R,T) - i\xi(R,T) \Big]D_I^>(t,R) 
\nonumber \\
&\equiv &
\Big[ 2m_Q+{\rm Re}\ V(R,T) - i{\rm Im}\ V(R,T) \Big]D_{\rm I}^>(t,R),
\eeq
with $V(R,T)$ being the proper heavy-quark potential at finite temperature.
On the other hand,
if there are no well-defined peaks in $\rho_{\rm I}(\bar{\omega},R)$,
 description of the $Q\bar{Q}$ system in terms of the potential is not justified.
 Note that the existence of  ${\rm Im}\ V(R,T)$ was first pointed out in \cite{Laine:2006ns}
  in which the late time behavior of $D^>(t,R)$ was calculated in hard
  thermal loop resummation  at high $T$.

\subsection{Thermal Wilson loop and its spectral decomposition}

To evaluate the spectral function defined in Eq.(\ref{eq:SPF-I})
 in non-perturbative lattice QCD simulations,
 we consider the imaginary-time correlator  
 $D(\tau,R) \equiv \left.
 \langle {\rm T}_{\tau} M_R(\tau) M_R^{\dagger}(0) \rangle 
 \right|_{0 \le \tau < 1/T} =  \langle  M_R(\tau) M_R^{\dagger}(0) \rangle $.
Since the heavy quarks are assumed not to propagate in the spatial directions, 
 their imaginary time propagator has the form
  $S_E(\mathbf{x}-\mathbf{x}',\tau-\tau')=S_E(\tau-\tau')\delta(\mathbf{x}-\mathbf{x}')$  
 with $S_E(\tau-\tau')$ being written in terms of temporal Wilson-lines.
For the full correlator we obtain
\beq
\label{eq:loop}
 D(\tau,R)=& & - A_{+-} \ e^{-2m_Q \tau}\ 
 {\rm Tr}[{\cal{W}}_{\rm I}(\tau,R)]  
 \qquad \qquad  \qquad \qquad  {\rm (loop)}\\
\label{eq:staple}
& &+A_{-+}\ e^{-2m_Q(\beta-\tau)}\ 
{\rm Tr}[{\cal{W}}_{\rm II}(\tau,R)]  \qquad  \qquad \qquad {\rm (staple)}\\
\label{eq:handle}
& & -A_{++}\ e^{-m_Q \beta }\ 
{\rm Tr}[{\cal{W}}_{{\rm III}a}(\tau,R)+{\cal{W}}_{{\rm III}b}(\tau,R)] 
 \qquad {\rm (handles)},
\eeq 
where the numerical coefficients are defined as e.g.
 $A_{+-}={\rm Tr}[\Gamma\Lambda^{(+)}\bar{\Gamma}\Lambda^{(-)}]$ with 
  $\Lambda^{(\pm)}$ being the projection operator onto the upper and lower components
 of the Dirac spinor.  ${\cal W}_{\rm I,II,III}(\tau,R)$ are the 
 Wilson-loops at finite temperature with different topological structure (loop,
 staple and handle) as shown in Fig.1. 

\begin{figure}[t]
\vspace{-0.4cm}
\begin{center}
\includegraphics[scale=0.43]{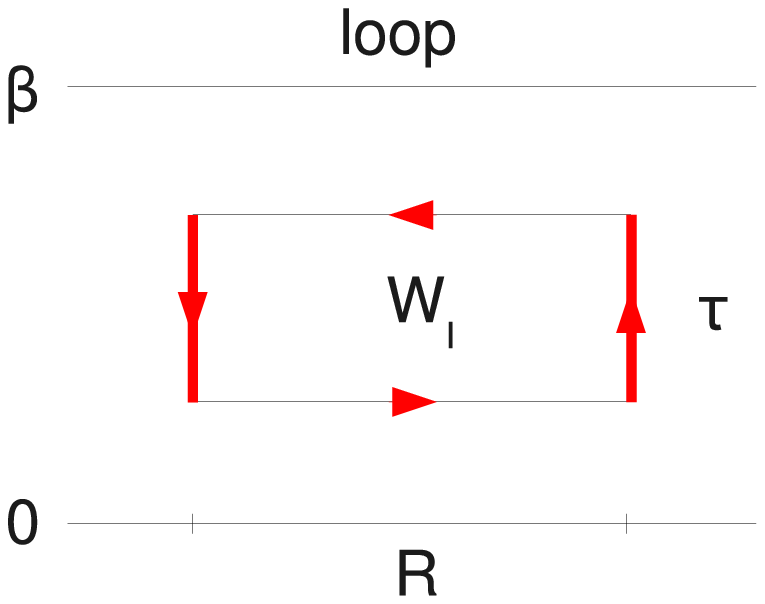}\ 
\hspace{0.1cm} \includegraphics[scale=0.43]{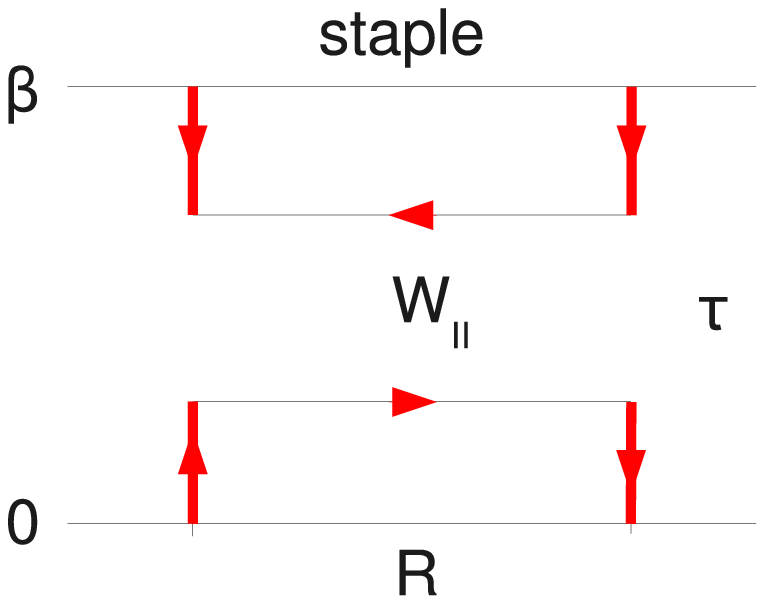}\ 
\hspace{0.1cm} \includegraphics[scale=0.43]{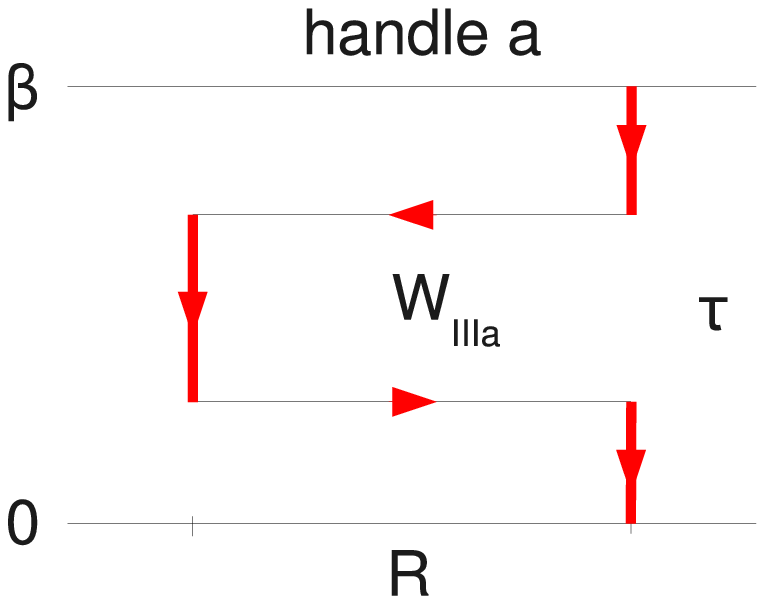}\ 
\hspace{0.1cm} \includegraphics[scale=0.43]{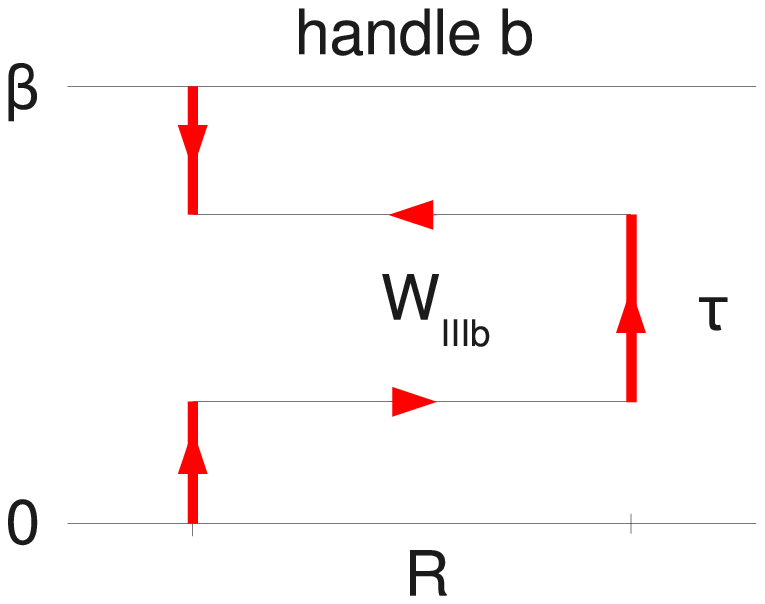}\ 
\end{center}
\vspace{-0.5cm}
\label{FigContours}
\caption{Wilson lines present in the determination of the imaginary-time 
$Q\bar{Q}$ correlator $D(\tau,R)$. The loop, staple and handles correspond to the 
 case I, II and III defined in Sec.2.2, respectively.}
\end{figure}

 From their $\tau$-dependence and particle content in the intermediate
 state, it is easy to see that loop, staple and handle correspond to the case I, case II
  and case III respectively.  Namely the loop contribution $D_{\rm I}(\tau,R)$ given in
  Eq.(\ref{eq:loop}) has the spectral decomposition,{ 
\beq
D_{\rm I}(\tau,R)  
= e^{-2m_Q \tau} \int_{-\infty}^{\infty}e^{-\bar{\omega} \tau} \rho_{\rm I}
(\bar{\omega},R) d\bar{\omega}.
\label{eq:DI-tau}
\eeq 
Thus we arrive at the following  formula relating the thermal Wilson loop
 with the spectral function,
\beq
{\rm Tr}[{\cal{W}}_{\rm I}(\tau,R)]
 =- \frac{1}{A_{+-}} \int_{-\infty}^{\infty}
	e^{-\bar{\omega}\tau}\rho_{\rm I}(\bar{\omega},R) d\bar{\omega}.
	 \label{eq:central}
\eeq
Note that this relation is well-defined in the  limit  $m_Q \rightarrow \infty$.

Lattice QCD simulations of the left hand side of Eq.(\ref{eq:central})
 for different values of $\tau$ and $R$
  allow a non-perturbative determination of the
  spectral function through the inverse Laplace transform,  e.g. by using 
  the MEM.
  The peak position and its width can consequently be translated into 
     the real and imaginary part of the proper potential.

\section{Quenched QCD results  at low and high $T$}

To study the feasibility of the method proposed above,
we perform quenched lattice QCD simulations  
 using  the plaquette gauge action with $\beta_{\rm lat}=7.0$ on 
 an anisotropic  $N_{\sigma}^3 \times N_{\tau}= 20^3 \times (96-32)$ lattice.
  The physical lattice spacing and anisotropy are the 
  same with the first reference in \cite{Asakawa:2003re}, 
   $a_\sigma=4a_\tau=0.039\rm fm$. 
  We adopt the fixed scale method \cite{Umeda:2008bd} 
where one varies $N_{\tau}$  to change the 
temperature.  An advantage of this  method is that the lattice spacing is the
 same for all temperatures so that 
  the  $Q\bar{Q}$ potentials for different temperatures can be
  directly compared without any adjustment.
   The thermal Wilson loop ${\cal W}_{\rm I}(\tau, R)$ 
 is calculated as a function of $\tau$ and $R$. We report  here only the 
results at the lowest temperature ($T=0.78T_c$) and
 the highest temperature ($T=2.33T_c$) using ($125,980$) gauge configurations respectively.
 
\begin{figure}[t]
\begin{center}
\hspace{-1cm}
 \includegraphics[angle=-90,scale=0.31]{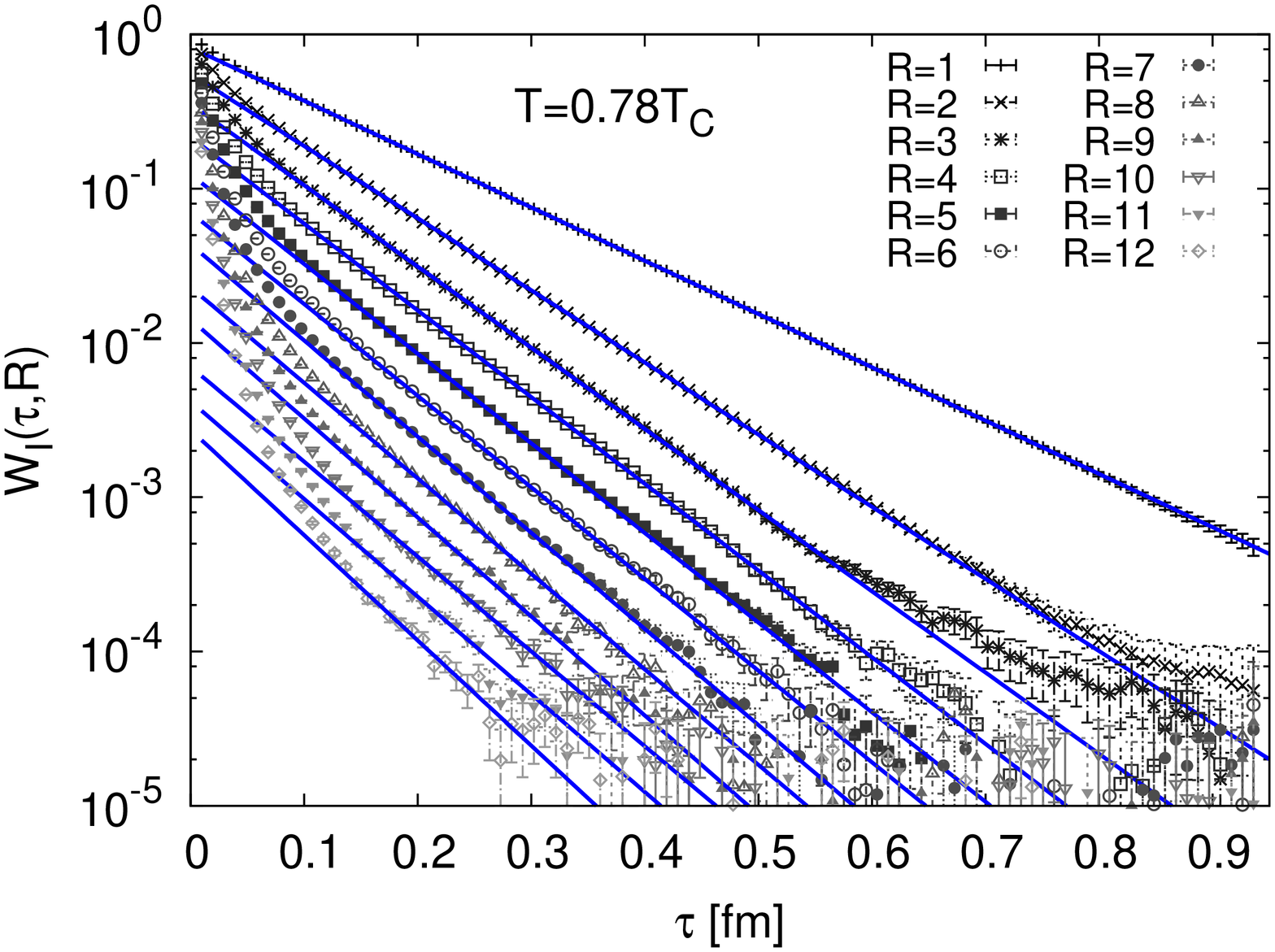}\ 
 \includegraphics[angle=-90,scale=0.31]{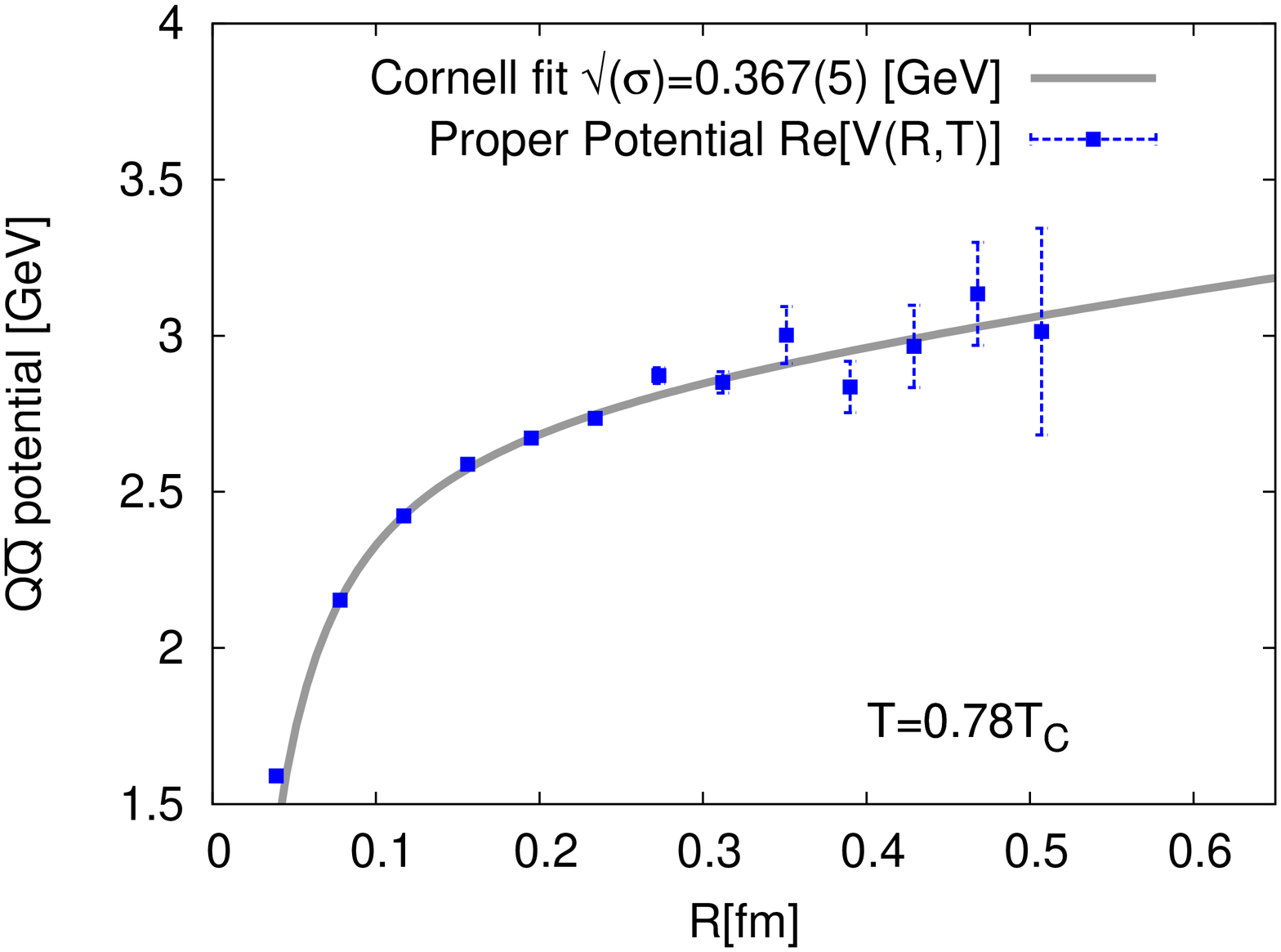}
\end{center}
\vspace{-0.3cm}
\label{FigBELOWT}
\caption{(Left) The thermal Wilson loop ${\cal W}_{\rm I}(\tau,R)$ as a function
 of $\tau$ at $T=0.78T_c$. The solid lines are the results of the 
 single exponential fit in the intermediate region of $\tau$.
  (Right) Proper potential reconstructed from the single exponential fit of the 
  thermal Wilson loop. The gray line
   here corresponds to a fit based on the Cornell-type potential
 $V(R)=c-\alpha/R+\sigma R$ with a string tension of 
 $\sqrt{\sigma}=367(5){\rm MeV}$ }
\vspace{-0.4cm}
\end{figure}

Shown in the left panel of Fig.2 is the on-axis thermal Wilson loop 
 ${\cal W}_{\rm I}$
  as a function of $\tau$ for different values of $R$ in the 
   low-temperature confinement phase ($T=0.78 T_c$).
    By definition, it is not symmetric
  under the reflection $\tau \leftrightarrow \beta-\tau$. In the small $\tau$ region,
  an effect from $Q\bar{Q}$ + excited gluons can be seen, which is  similar to the 
 case of the standard hadronic correlation functions.  In the intermediate
  $\tau$ region, we find a single exponential behavior, which  
   suggests the existence of a distinct peak of vanishing width in the spectral function. The position of the 
  peak is nothing but the real part of the proper potential $V(R,T)$.
  We make a single exponential fit of  ${\cal W}_{\rm I}$ for each $R$ in the 
  interval ($ 15a_\tau \lesssim \tau \lesssim 30 a_\tau $).  The result is plotted as a function of $R$
   in the right panel of  Fig.2 with filled squares. 
  Error bars estimated by a $\chi^2$ fit for several slightly shifted or contracted fitting regions
  reflect both the statistical and systematic uncertainties. 
  The data  can be fitted well by a Coulomb + linear form with
  an effective string tension $\sqrt{\sigma_{_{T=0.78T_c}}} = 367 (5) {\rm MeV}$
   as indicated by the gray line.
  This is smaller than the known value at $T=0$,  $\sqrt{\sigma_{_{T=0}}}
   \simeq 430 {\rm MeV}$
  possibly due to the thermal fluctuations of the confining string. 
   We emphasize here that
   (i) the thermal Wilson loop is directly linked 
    to the Schr\"{o}dinger equation as we have discussed  and  
   (ii)  one can obtain not only the
   real part but also the imaginary part of the proper potential from the 
    thermal Wilson loop by using e.g. MEM.
   Analysis of the present lattice data in the whole range of $\tau$
    by using  MEM  is currently under way.

\begin{figure}[t]
\vspace{-0.3cm}
\begin{center}
\hspace{-1cm}
 \includegraphics[angle=-90,scale=0.31]{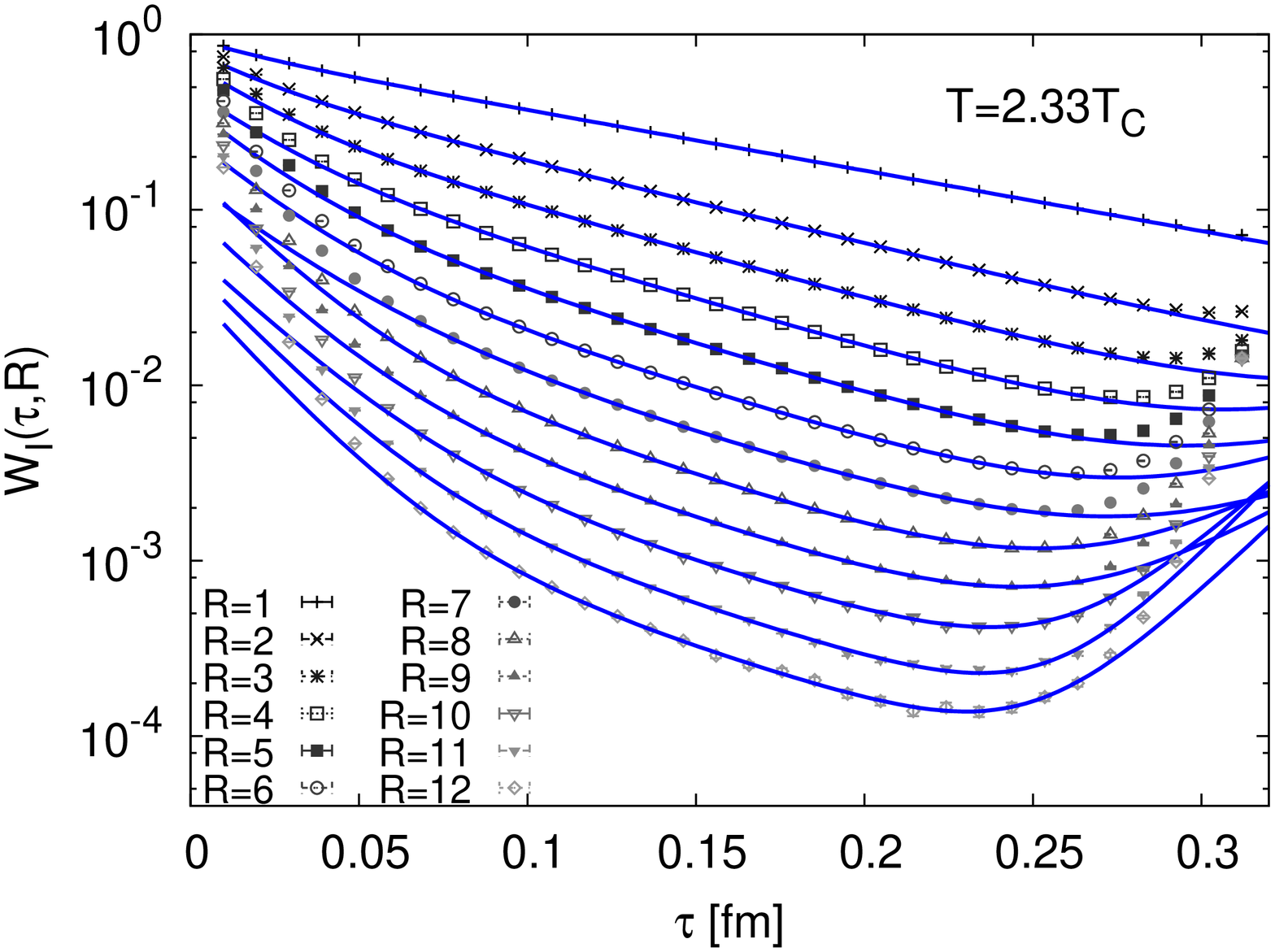}\ 
 \includegraphics[angle=-90,scale=0.31]{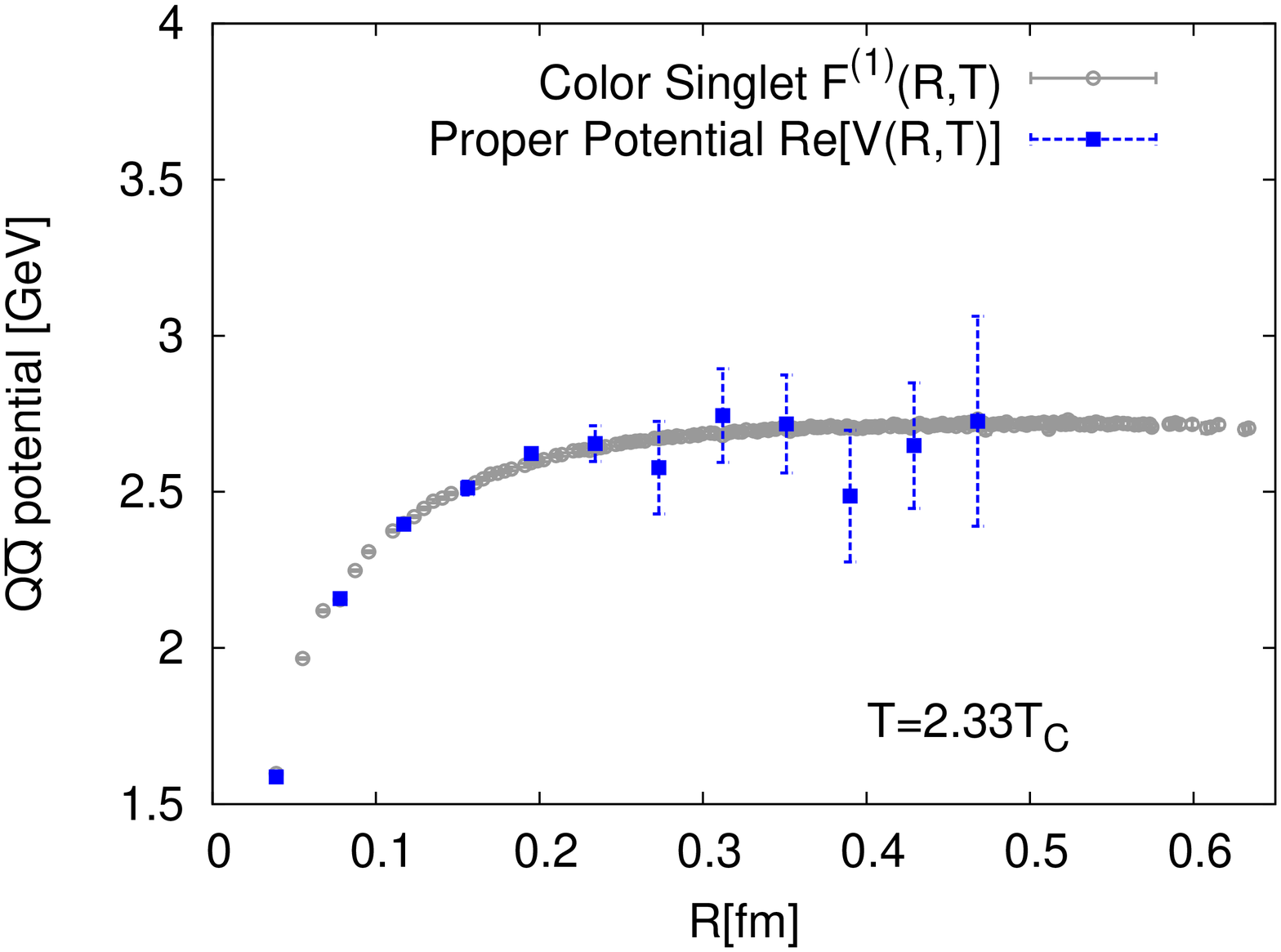}
\end{center}
\label{FigABOVET}
\vspace{-0.4cm}
\caption{(Left)  The thermal Wilson loop ${\cal W}_{\rm I}(\tau,R)$ as a function
 of $\tau$ at $T=2.33T_c$. The solid lines in the left panel
 are the results of the 
 triple exponential fit in the intermediate region of $\tau$. (Right) Proper potential reconstructed from the triple exponential fit of the 
  thermal Wilson loop. Color-singlet potential obtained from the Polyakov-line
   correlation in the color-singlet channel is also shown for comparison.}
\vspace{-0.2cm}
\end{figure}

 Shown in the left panel of Fig.3 is the thermal Wilson loop 
 ${\cal W}_{\rm I}$ in the high temperature deconfinement phase ($T=2.33T_c$). 
  In this case, we find three characteristic features:  For 
  $\tau/a_{\tau} \ll N_{\tau}$, some 
  evidence for excited states can be seen.
   For  $\tau/a_{\tau} \sim N_{\tau}/2$, there appears an approximate
    single exponential behavior
    which  indicates a well-defined peak in the spectral function.
    For  $\tau/a_{\tau} \sim N_{\tau}$, the thermal Wilson loop increases
    as $\tau$ increases, which  indicates that 
    the spectral function $\rho_{\rm I}(\bar{\omega},R)$  has some structure
    for negative $\bar{\omega}$ although
     the spectral strength is extremely small.  This may be interpreted 
    as the effect of thermal gluons with energy
     $\epsilon_{\rm th}$; they can compensate a deficit of the 
     external energy $\bar{\omega}$
      to match the energy conservation, $2m_Q = \omega +   \epsilon_{\rm th}$ or equivalently $\bar{\omega}=-\epsilon_{\rm th}$,
       so that a $Q\bar{Q}$ can appear in the intermediate  state.  
  A triple exponential fit is deployed, designed to give stable results in the presence of
  non single-exponential behavior coming from the end regions of the $\tau$ interval. 
 The fit results of  ${\cal W}_{\rm I}$ for each $R$ in the 
 interval ($ 10a_\tau \lesssim \tau \lesssim 20 a_\tau $) leads to the plot
   in the right panel of  Fig.3. 
   It shows a Coulombic behavior at short distances, while the potential
   seems to be Debye screened  at long distances.   For comparison, we measure 
    the color-singlet free energy  on the same lattice from  
   the Polyakov line correlations in the color-singlet channel under Coulomb
     gauge fixing: 
 $ F^{(1)}(R,T) = -T \ln \langle {\rm Tr} \Omega({\bf x}) \Omega^{\dagger}({\bf y}) \rangle$.
   The results are shown by the open circles.
  Although ${\rm Re} \ V(R,T)$ and $F^{(1)}(R,T)$ have no direct theoretical
   connection, they coincide within the error margins devoid of any adjustments. 
   An analysis of the present lattice data over the whole range of $\tau$ by using MEM
    is currently under way to extract the complete spectral structure of $\rho_{\rm I}$.  
    This will enable us to 
    extract   ${\rm Im} \ V(R,T)$ and also to judge the validity of the 
     potential picture at high $T$. 

\vspace{-0.2cm}
\section{Summary and concluding remarks}
\vspace{-0.2cm}

We proposed a non-perturbative and gauge invariant approach to connect the
 Sch\"{o}dinger equation description
of a heavy $Q\bar{Q}$ pair in terms of a static potential with the spectral function of the 
thermal Wilson loop obtained from Lattice QCD. It was shown that if the spectral structure 
is well defined for a set of temperatures $(T)$ and separation distances $(R)$, the peak position and 
width correspond to the real and imaginary part of the proper potential $V(R,T)$ respectively. 
A first determination of the real part of the proper potential 
from quenched lattice QCD simulations ($N^3_\sigma \times N_{\tau} =20^2 \times (96-32)$) was presented for $0.78T_c$ and $2.33T_c$.
It showed that although no apparent connection between the real part of the proper potential and the color-singlet free-energy
potential exist, their values above $T_c$ coincide
 within the error bars.
We are currently analyzing the thermal Wilson loop data for 
$0.78 \le T/T_c \le 2.33$, using MEM to determine the
full spectral structure in an attempt to reconstruct not only the real but also the imaginary part,
especially above the phase transition. In the case of infinitely heavy quarks, the only contribution to such an imaginary part at high $T$ comes from the scattering of light medium particles, i.e. Landau damping \cite{Laine:2006ns,Beraudo:2007ky}. The non-perturbative
 determination of  ${\rm Re}\ V(R,T)$ and ${\rm Im}\ V(R,T)$ for $1 \le T/T_c \le 2$ is
  particularly important in relation to the fate of charmonium above $T_c$.

Two directions for future research are thus in order: (i) we need to include the effects of light fermions
in the medium by utilizing full QCD $N_f=2+1$ configurations, and (ii) the approach has to be extended to finite $Q\bar{Q}$ masses. As for (ii),
a similar strategy as for $T=0$, where the spatial fluctuations 
of the heavy quarks are incorporated by appropriate insertion of handles in the temporal Wilson lines \cite{Koma:2006si} 
might be a first starting point.

\vspace{0.3cm}

 We thank Yuu Maezawa and the members of the WHOT-QCD Collaboration for
 useful discussions.  
 This research was supported in part by the Grant-in-Aid of MEXT (Nos. 
18540253) and by Grant-in-Aid for Scientific Research on Innovative
Areas (No. 2004: 20105003).

\end{document}